# Vast Educational Mobile Content Broadcasting using ARMrayan Multimedia Mobile CMS

Somayeh Fatahi

Department of Computer Engineering
Kermanshah University of Technology
Kermanshah, Iran
fatahi_somayeh@yahoo.com

Ali Reza Manashty and Zahra Forootan Jahromi

Department of Computer Engineering
Razi University
Kermanshah, Iran
a.r.manashty@gmail.com, zahra.forootan@gmail.com

*Abstract*— The huge information flow currently available to young generation makes it difficult for educational centers to train them as needed. Most of these information flows occur in transportation time or while on public areas. Competing with commercial information streams is far out of educational centers time and budget. For creating enough mobile applications for vast educational mobile content broadcasting that can match young spirits as well, we designed and developed the ARMrayan Multimedia Mobile CMS as the software that helps communities, educational, cultural or marketing centers in a way that ordinary operators be able to create a variety of fully functional multimedia mobile applications such as tutorials, catalogues, books, and guides in minutes without writing even a line of code. In this paper, we present the role of our developed software in our proposed vast educational content broadcasting system using kiosks and Bluetooth advertising, which will lead to a great leap in M-commerce marketing and public education solutions. Related experiences are described and diagrams are used to illustrate the solution. Upon release of the software, it achieved two titles and prizes in different festivals and various cultural and commercial centers became its customers.

*Keywords- mobile education; m-commerce; mobile CMS; multimedia cms; mobile; content broadcasting; mobile catalogue; education; J2ME.*

## I. INTRODUCTION

Nowadays, the use of cell phones as an everywhere company has become a habit for nearly everyone. E-Commerce, which is based on using electronic devices for business uses electronic marketing as the main means of advertising. While e-commerce mostly relies on web and internet, an area which focuses mostly on marketing and trading using the mobile phones of the consumers, is called Mobile Commerce (M-Commerce). This area of commerce mostly involves contacting users via their mobile phones. Recent business models and applications of M-Commerce are fully described by Chen Xin [1]. This includes using mobile networks to send text/multimedia contents such as SMS and MMS to proximity marketing which is based on sending mobile contents using Bluetooth technology to the cell phones in the range of the Bluetooth sender device.

The huge information flow currently available to young generation makes it difficult for educational centers to train them as needed. Most of these information flows occur in transportation time or while on public areas. Competing with commercial information streams is far out of educational centers time and budget. For creating enough mobile applications for vast educational mobile content broadcasting that can match young spirits as well, we designed and developed the ARMrayan Multimedia Mobile CMS as the software that helps communities, educational, cultural or marketing centers in a way that ordinary operators be able to create a variety of fully functional multimedia mobile applications such as tutorials, catalogues, books, and guides in minutes without writing even a line of code.

Public education affects many areas. First it is closely in contact with industry as both affect each other in budget and policy [2]. In long term, public education also helps removing the inequality between poor and rich [3] and help communities improve education quality [4], so the best practice for improving justice in society education is investing in innovative public education methods.

In proximity marketing, as mentioned earlier, the contents that can be sent using the Bluetooth devices vary from virtual cards (VC) to text and multimedia contents such as sound, video and picture and even mobile applications. It is nearly impossible to send various different contents such as a text and picture using for example a single HTML file, because most of the users only accept the incoming connection once from each Bluetooth advertising devices and most possibly will be out of range of the device (lose connection) after the first file received. The later problem is a permanent problem because of the limited range of Bluetooth 2.0 EDR technology for mobile phones which is at most about 150 meters.

The problem of sending various related contents using proximity marketing was a challenge which was resolved with the use of mobile E-book makers, especially those with the capability of displaying Unicode language of the region regardless of the target mobile device support. The output of such mobile E-book makers is a stand-alone J2ME mobile application which can be sent as a single .jar file and is installable on most of the mobile phones that support Java platform.

Most of the primary mobile e-book makers, especially in Iran, which could support Farsi language, such as Parnian Mobile E-book Studio, could only support text and a theme. The next generation of these E-book makers could support pictures in the text too. Still none of them has the capability of supporting rich media contents such as sounds and videos.

Nowadays people spend a notable amount of time in transportation, without having access to their PCs or having hard time accessing their laptops; instead, they are constantly using their cell-phones/PDAs. Because of this, many companies and active advertisement centers, whether religious, educational or business related are pinpointing on these handset devices for applying their policies. Despite of this huge trend, there is still a lack of programmers that write both visually and functionally acceptable mobile applications in beneficial time and cost.

Due to the lack of rich media support in such content-oriented mobile applications and the need for fast production of such mobile applications, we designed and implemented ARMrayan Mobile Multimedia CMS (Content Management System) [5] to omit the programming part of producing content oriented multimedia mobile software. This software help communities, educational, cultural and marketing centers in a way that ordinary operators be able to create a variety of fully functional multimedia mobile applications such as tutorials, catalogues, books, and guides in minutes without writing even a line of code.

In this paper, we present the role of our developed software in vast educational proximity marketing systems using Bluetooth advertising, which will lead to a great leap in M-commerce marketing solutions. Educational and cultural rich content mobile applications can now be produced easily and distributed using proximity marketing technologies, reaching the goal for a higher educated and cultural society. Related experiences are described and diagrams are used to illustrate the solution that involves a two level updating procedure for kiosks that involve end-users using proximity marketing features.

## II. PROPOSED METHOD

As mentioned earlier, people waste a lot of time in transportation. In big cities, this transportation might take more than an hour. There has been many commercial products which involve taking advantages of these wasted times of people with teaching skills they mostly need to learn, such as language and psychological skills. Still there is lack of governmental effort to take advantage of these wasted times. Non-electronic cultural media such as small books and newspaper is not that suitable because of their volume and free-space consumption. A heavy bulk of newspapers and books will tire even the strongest people. Here is where the important point emerges: Mobile phones that are carried by all people every day.

### A. Importance

By end-2008, there were nearly 3.9 billion mobile subscribers worldwide and it is expected that in 2013 this number will reach close to 6 billion [6]. The fact above, shows that even people are not eager to carry books and paper advertisement everywhere, mostly in transportation, they will not leave behind their mobile phones. Users' mobile phones are the greatest nest for cultural contents and commercial ads. With operators mobilizing to explore and capture potential markets in the developing nations, it is expected that Asia Pacific will house nearly 50 percent of the worldwide mobile subscriber base by end-2013 [6]. This also means that the continent of culture and tradition will be the host of half of the cell phone carriers in near future. Governments of Asian countries are the most possible demanders of transferring their traditional to the new generation to keep it more alive than ever.

### B. ARMrayan Multimedia Mobile CMS

Now that mobile devices are non-detachable part of everybody's routine; there is still a need for mobile content creation tools so that the providers of contents can publish their data (whether rich-content data or simple texts) through structured theme-based applications that can also provide some facilities for using mobile device special capabilities (e.g., sending contents through SMS/MMS or media playback; that using mobile browsers for these jobs is still facing some difficulties).

In web-based services, the term CMS (Content Management System) is used to describe a web application that gives content-oriented accessibility to the administrator of the website, regardless of his/her knowledge in programming concepts of the website. In addition to content-oriented access, the administrator can also change the visual theme of his/her website. There's still a need for a system that provides tools for easy management of contents for use in mobile application providing additional mobile device facilities for efficient distributing or viewing of mobile-based contents. Such applications may now be called Mobile Content Management Systems or MCMSs.

The ARMrayan Multimedia MCMS is a PC-based application providing the user the tools needed for inserting his/her desired multimedia content in a tree-based index of pages, in addition to providing simple theme-based UI designing facilities. The role of our implemented software is to simplify the content transferring process by creating a means for producing rich media content-oriented mobile software. The output of this Windows application is a J2ME mobile application with the desired content and visual design. The J2ME application was developed separately using the Java language for MIDP2.0+ supporting mobile phones. This base J2ME file that the Windows program receives as input is called the Mobile Engine File. This Engine File can be modified or updated regularly to maintain compatibility with newer devices and/or get improvement in the UI or search engine features.

*1) Implementation and Usage*

For using ARMrayan MCMS, an operator, managed by the system administrator, designs a tree index of pages in the Windows program and then he/she will now be able to add several contents per page in the desired order. These content types can currently be among texts, pictures, sounds or music, videos and animations, map points, phone number, e-mail address, web sites link and etc. It is also possible to change the order of the contents within each page. The colors, background pictures and the background music of different areas of the output mobile application can also be set in the main user interface.

Implementation of the Windows application was using Visual C# .Net while the mobile engine application was developed in Java 2 Micro Edition language. The main features of the implemented software are as following:

- Supporting texts, sounds, images, videos, phone numbers, web links and emails as the input content.

- Tree-view page designing.

- Displaying Middle Eastern fonts via the bitmap font framework.

- Multilanguage environment (English and Farsi)

- Environment theme designing.

- Sending contents via SMS/MMS.

- Text searching.

*2) Preview*

The main program engine displays the content in tree views thus making it possible for a page to have innumerous subpages (Fig. 1). The mobile engine of the ARMrayan MCMS was developed as J2ME application. This engine is actually a stand-alone mobile application simply known as a "Mobile Engine" in the application, containing the main algorithms for displaying or playing different contents such as sounds, videos and texts. This mobile software (Fig. 2) also contains the functions to show the menus and performing other operations such as searching within the content and handling the bitmap font for Asian languages.

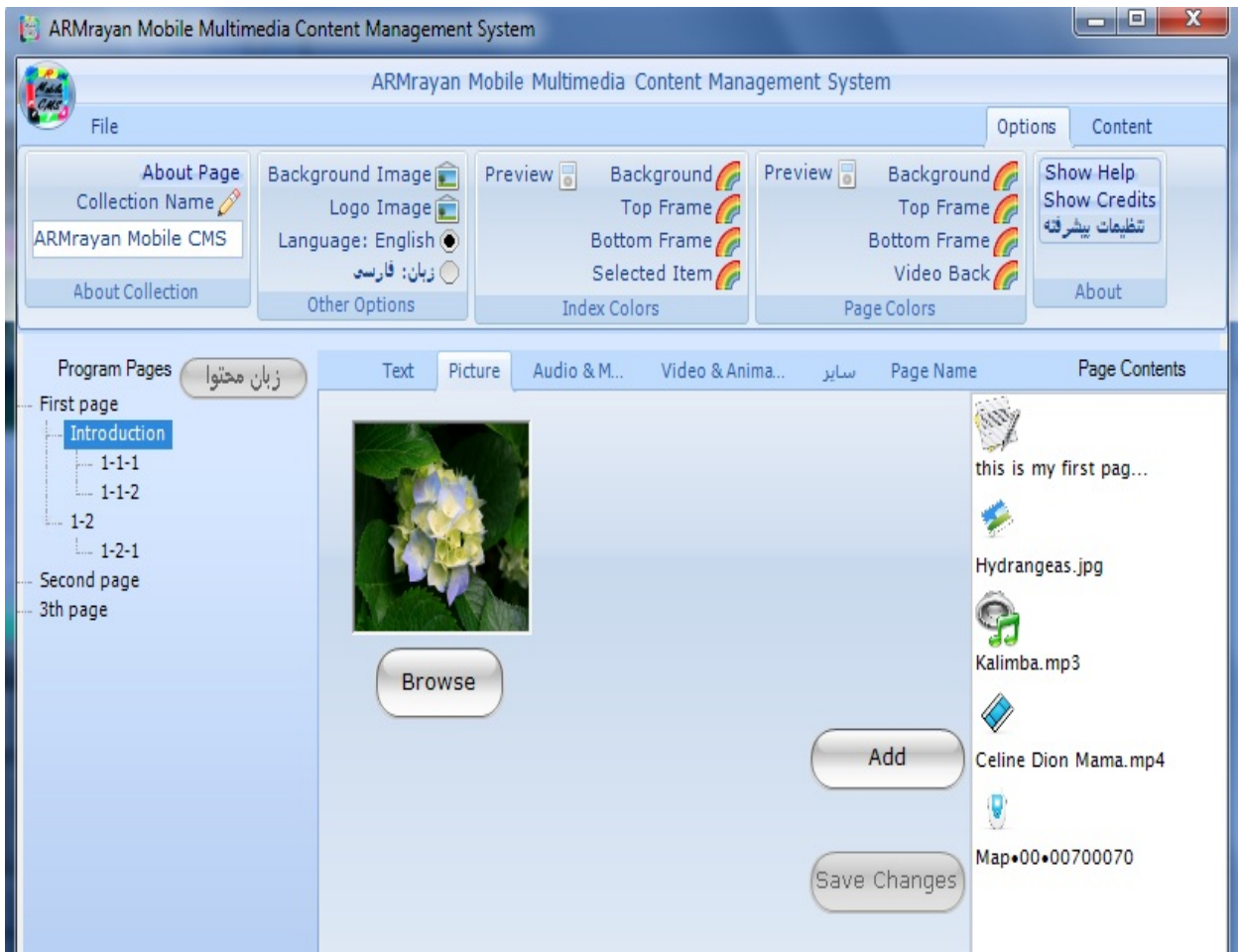

Figure 1. ARMrayan Multimedia Mobile CMS, Windows interface for inserting multimedia contents and designing final mobile application for creating final mobile application

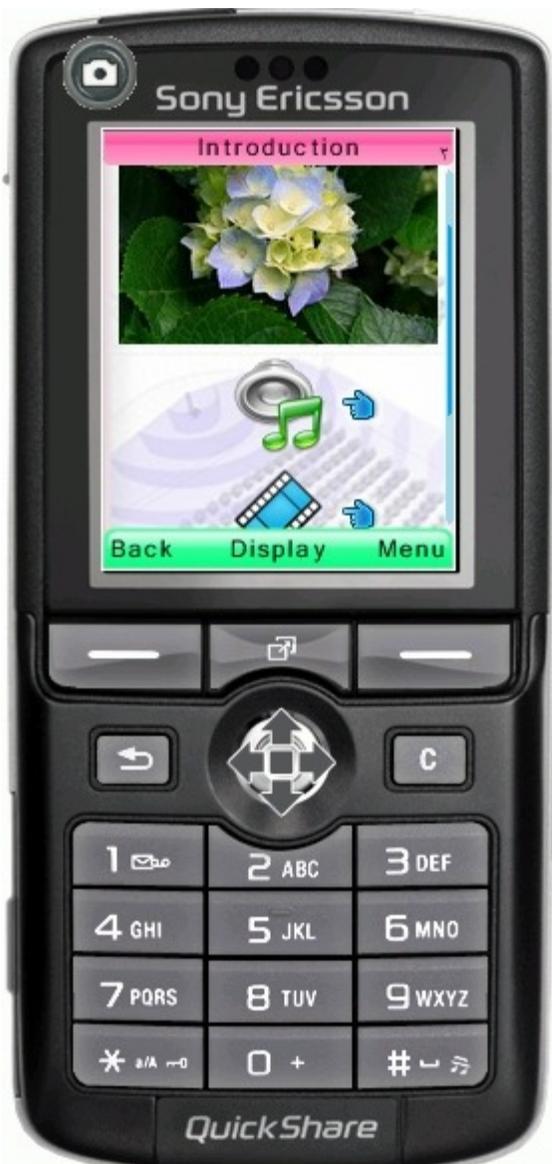

Figure 2. The output Java mobile application resulted from the design and contents of ARMrayan MCMS illustrated in (Fig. 1).

## C. *Vast Educational Mobile Content Broadcasting*

It was described earlier that the most vital part of each society that can be the within the range of electronic cultural and educational goods, are the people using cell phones every day, mostly in transportations. This can be a target for both educational and cultural centers to broadcast their content to the society. This is done by gathering contents from certified sources and encapsulating the contents in a well-designed mobile application. Now we describe our proposed system for broadcasting vast educational multimedia mobile contents that is produced using ARMrayan Multimedia Mobile CMS and will be delivered using Proximity Marketing Systems.

### 1) *Creating and Broadcasting Contents*

An operator, under the supervision of an administrator, inserts multimedia contents to the ARMrayan Multimedia

MCMS and receives a Java mobile application (a .jar file) as the output. While categorizing the content-based applications for broadcasting in several servers, the operator uploads the file to a central server. Various other sub-servers collect their appropriate categorized applications, adding new ones or updating current mobile applications, and start updating the several Proximity Marketing Systems that are in contact with the end-users (people) (Fig.3).

### 2) *Sub-Servers*

The various sub-servers that are designed to broadcast a series of categorized applications can be connected to multiple Proximity Marketing Systems (such as kiosks) in different places of a metropolitan area. After being updated from the main server, the sub-servers can disconnect from the main server and start updating the kiosks under their control. The category and kiosk locations can vary by the end-user groups and content intended to broadcast. For example a commercial advertisement group may use a dedicated sub-server to broadcast advertisement contents in various large malls and shopping areas. On the other hand, a university can publish its latest discoveries and research results along with educational contents and tests using a server while other governmental cultural centers can place such kiosks in public transportation areas where people can automatically or manually receive educational mobile e-books with rich multimedia contents that are easily readable and portable within transportation stations and along the day.

### 3) *Proximity Marketing Systems*

The Proximity Marketing Systems used in this project are some kiosks that have embedded Bluetooth devices that can function in two different ways:

- Automatically broadcasting mobile application software that is designed to be shared to general public using Bluetooth technology within the range of the device.

- Manually receivable mobile applications that is chosen from the inner menu of the kiosk stations. This offers a wider variety of mobile applications to be shared for specific persons that intend to find their suitable content themselves.

## III. CONCLUSION AND RESULTS

We've tested a Proximity Marketing System for one of our mobile application files in the 3rd International Digital Media Festival and Exhibition, Tehran, October 2009. There was one server that was automatically sending the file to the customers. In each Bluetooth search the device was making, it could discover about 180 devices but most of them were vanishing too fast. The device could only send 7 files simultaneously within the range of about 100 meters in radius. In a 2-day test, a total of about 1800 attempt to send the file has been made by the device, from which about 600 were successful, 200 failed and 1000 were rejected by the customers. During this time, no more than about 100 visits were made to our place, where some of which found us using

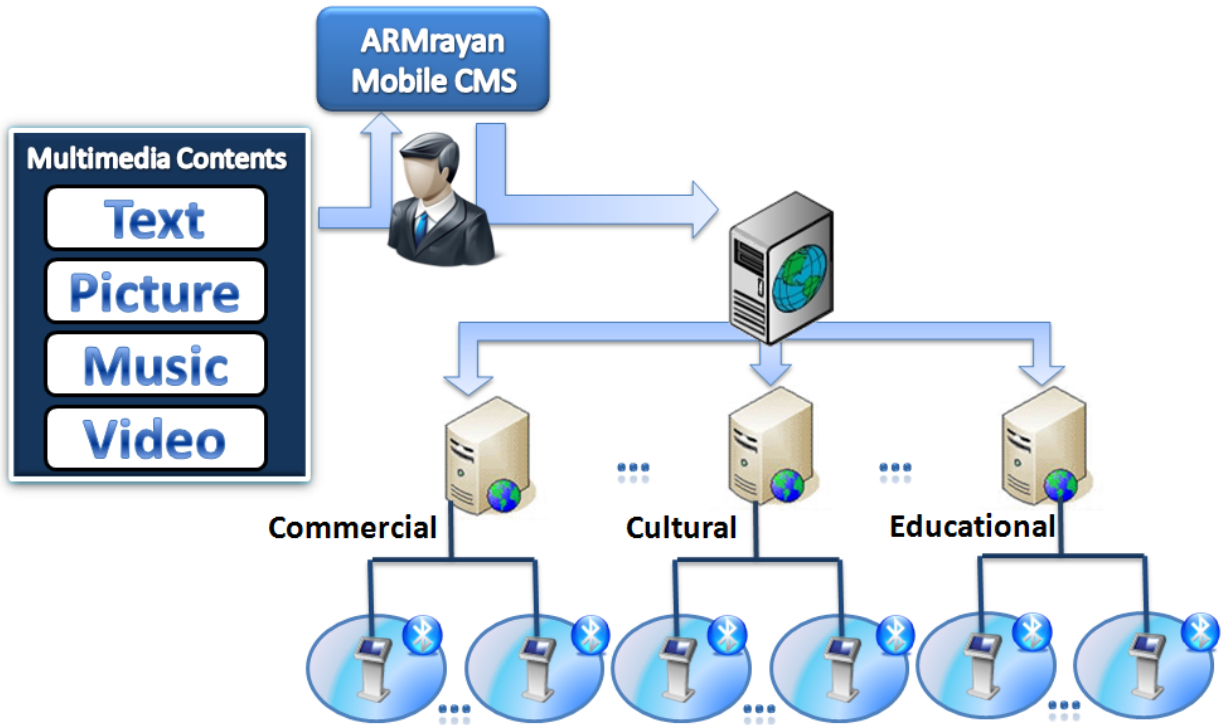

Figure 3. The proposed system diagram of Public Education and M-Commerce via Proximity Marketing Systems using ARMrayan Multimedia Mobile CMS.

the file we sent them earlier! It can be inferred from this attempt that:

- About 6 times more than the people that physically visited the commercial place, the people received the advertising information electronically using the Proximity Marketing System.

- More than 50% of the users rejected the file transfer, which means there need to be a huge effort for creating a safe and informative atmosphere so that people accept such files more easily and safely.

Such an experience repeated with just the same results, but with a decrease in rejections in the tourism activities in Norouz 89 (Starting of the New Year in spring). The reason of the reduced rejects was the banners informing the tourists to turn their Bluetooth on and accept the incoming file to inform them better about the travel they would made.

The multi-server based project described in this paper was proposed to the Kermanshah City government to be executed in several public transportation stations.

In the first month of developing this system, we awarded the 1st Place in the 3rd International Digital Media Festival as the "The Best Mobile Software in Technology, Innovation and Development" for "ARMrayan Mobile CMS"; Iran-Tehran; October 2009 [7]. Not many days passed, that an output of the program which contained audio and photo gallery, multi-language text and fabulous design, also awarded the 3rd Place in the 4th National Imam-Reza Festival, Professional Mobile Software Title of the Digital Media section, for the "Imam Reza Pilgrim Mobile Software"; Iran-Tehran; December 2009 [8]. We also sponsored by a high cultural and religion center for upgrading the application with their desired need, so that they can produce multi-language cultural and religious content-oriented mobile application for broadcasting into the society.

## IV. FUTURE WORKS

Our software is Windows-based and operator should upload the mobile application output file to the server upon creating it. It is considered to design and implement the ARMrayan Multimedia Mobile CMS as a web application in near future, so that the operator can create specific mobile application directly on the web by using the web based version of the software.

The other work that can be done in future is finding a way that users of our mobile applications are able to update their applications, which have installed on their phone, via using GPRS capability of mobile phones and get the new version of our different mobile applications. This is a very good solution for our users to be able to update their installed mobile applications without referring to the kiosks which are installed in a location that they might not be able to access it directly. A similar research but in another area of M-Commerce was done by Fumin Zou Shuling Zhang [9], where he presented a model for internet system for M-Commerce in trains.

Although some research has been done about the factors affecting people in mobile advertising from mobile operators [10, 11, 12], more research on the following area is considered helpful, with consideration of people's opinion in several places:

- What are the effects of the banners on people's acceptance or rejection of the file?

- What type of file do people mostly prefer to accept?

- What is the effective area of each Proximity Marketing System?

- What is the efficient size of the file to send to the people?

- Receiving the feedback of people about the contents they received and the one they desire the most.

AUTHORS PROFILE

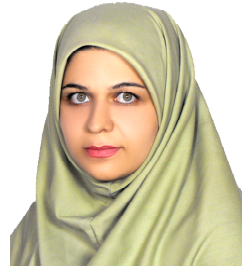

**Somayeh Fatahi** received her B.Sc. and M.Sc. degrees in Computer Engineering from Razi University, and Isfahan University, Iran, in 2006 and 2008 respectively and is now a Ph.D. student in University of Tehran, Iran. She is Lecturer in the Department of Computer Engineering at Kermanshah University of Technology. She also teaches in Razi University, University of Applied Science and Technology, Payam Noor University of Kermanshah, Islamic Azad University of Kermanshah, Institute of Higher Education of Kermanshah Jahad-Daneshgahi. Her research activities include (1) simulation of agents with dynamic personality and emotions (2) Computational cognitive modeling (3) Simulation and formalization of cognitive processes (4) Multi Agent Systems (5) Modeling of Human Behavior (6) Fuzzy Expert Systems.

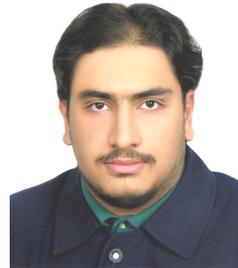

**Ali Reza Manashty** is a senior B.S. student in Software Computer Engineering at Razi University, Kermanshah, Iran and will be graduated in September 2010 awarding the 3rd rank among the autumn 2006 entrance students of Computer Engineering Department. He is going to continue his academic career as a M.Sc. student in the next semester. He has been researching on mobile application design and smart environments especially smart digital houses since 2009. His publications include 4 papers in international journals and conferences and one national conference paper. He has earned several national and international awards regarding mobile applications developed by him or under his supervision and registered 4 national patents. He is a member of Exceptional Talented Students office of Razi University since 2008 and he was the teacher assistant of several under-graduate courses since 2008.

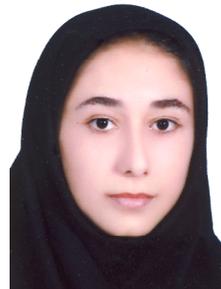

**Zahra Forootan Jahromi** is a senior B.S. student in Software Computer Engineering at Razi University, Kermanshah, Iran and will be graduated in September 2010. She is now researching in smart environments specially on simulating smart digital homes. Her publications include 4 papers in international journals and conferences and one national conference paper. She has 3 registered national patents and is now teaching Robocop robot designing for elementary and high school students at Alvand guidance school. She is a member of Exceptional Talented Students office of Razi University since 2008.